\documentclass[twocolumn,prl,aps,superscriptaddress, longbibliography, ]{revtex4-1}
\usepackage[T1]{fontenc}
\usepackage[latin9]{inputenc}

\usepackage{amsmath}
\usepackage{amssymb}
\usepackage{xcolor}
\usepackage{bbm}
\usepackage{braket}
\usepackage{mathrsfs}
\usepackage{pdfpages}
\allowdisplaybreaks

\usepackage{comment}

\usepackage{graphicx}

\usepackage[colorlinks=true]{hyperref}  
\hypersetup{
    bookmarks=true,         
    unicode=false,          
    pdftoolbar=true,        
    pdfmenubar=true,        
    pdffitwindow=false,     
    pdfstartview={FitH},    
    pdftitle={Euler Insulators},    
    pdfauthor={Bouhon and Slager},     
    pdfsubject={},   
    pdfcreator={},   
    pdfproducer={}, 
    pdfkeywords={} {} {}, 
    pdfnewwindow=true,      
    colorlinks=true,       
    linkcolor=blue, 
    citecolor=blue,        
    filecolor=magenta,      
    urlcolor=blue,           
	breaklinks=true
}

\definecolor{orange}{rgb}{1,0.5,0}

\newcommand{\sect}[1]{\vspace{0.3em}{\it #1.}---}

\usepackage{bbold}

\newcommand{\be}{\begin{equation}}
\newcommand{\ee}{\end{equation}}
\newcommand{\bea}{\begin{eqnarray}}
\newcommand{\eea}{\end{eqnarray}}

\renewcommand{\vec}[1]{\boldsymbol{#1}}

\def\bs#1{\boldsymbol{#1}}
\def\dd{\mathrm{d}}

\newcommand{\ie}{{\it i.e.}~}

\makeatletter
\AtBeginDocument{\let\LS@rot\@undefined}
\makeatother

\begin{document}

\title{Observation of an acoustic topological Euler insulator with meronic waves}

\author{Bin Jiang}
\altaffiliation{These authors contributed equally to this work}
\address{Suzhou Institute for Advanced Research, University of Science and Technology of China, Suzhou 215123, China}
\address{School of Physical Science and Technology \& Collaborative Innovation Center of Suzhou Nano Science and Technology, Soochow University, Suzhou 215006, China}
\author{Adrien Bouhon}
\email{Corresponding author: adrien.bouhon@su.se}
\altaffiliation{These authors contributed equally to this work}
\address{TCM Group, Cavendish Laboratory, University of Cambridge, J. J. Thomson Avenue, Cambridge CB3 0HE, United Kingdom}
\affiliation{Nordita, Stockholm University and KTH Royal Institute of Technology, Hannes Alfv{\'e}ns v{\"a}g 12, Stockholm SE-106 91, Sweden}
\author{Shi-Qiao Wu}
\altaffiliation{These authors contributed equally to this work}
\author{Ze-Lin Kong}
\author{Zhi-Kang Lin}
\address{School of Physical Science and Technology \& Collaborative Innovation Center of Suzhou Nano Science and Technology, Soochow University, Suzhou 215006, China}
\author{Robert-Jan Slager}
\email{Corresponding author: rjs269@cam.ac.uk}
\altaffiliation{These authors contributed equally to this work}
\address{TCM Group, Cavendish Laboratory, University of Cambridge, J. J. Thomson Avenue, Cambridge CB3 0HE, United Kingdom}
\author{Jian-Hua Jiang}
\email{Corresponding author: jianhuajiang@suda.edu.cn}
\address{Suzhou Institute for Advanced Research, University of Science and Technology of China, Suzhou 215123, China}
\address{School of Physical Science and Technology \& Collaborative Innovation Center of Suzhou Nano Science and Technology, Soochow University, Suzhou 215006, China}

\date{\today}

\begin{abstract}
Topological band theory has conventionally been concerned with the topology of bands around a single gap. Only recently non-Abelian {topologies that thrive on involving multiple gaps} were studied, unveiling a new horizon {in topological physics} beyond the conventional paradigm. Here, we report on the first experimental realization of a topological Euler insulator phase with unique meronic characterization in an acoustic metamaterial. We demonstrate that this topological phase has several nontrivial features: First, the system cannot be {described} by  conventional topological band theory, but has a nontrivial Euler class that captures the unconventional geometry {of the Bloch} bands {in the Brillouin zone}. Second, we uncover in theory and probe in experiments a meronic configuration of the bulk Bloch states for the first time. Third, using a detailed symmetry {analysis}, we show that the topological Euler insulator evolves from {a non-Abelian topological semimetal phase via the annihilation of Dirac points in pairs in one of the band gaps}. With these nontrivial properties, we establish concretely an unconventional bulk-edge correspondence which is confirmed by directly measuring the edge states via {pump-probe techniques}. Our work thus unveils a nontrivial topological Euler insulator phase with {a unique} meronic {pattern} and paves the way as a platform for {non-Abelian topological} phenomena.
\end{abstract}
\maketitle 

\sect{Introduction}
Topological phases of matter~\cite{Rmp1,Rmp2} offer an intriguing realm with rich emergent phenomena that are unavailable in other regimes. Although the past decade has witnessed remarkable progresses in the study of topological phases~\cite{clas1,clas2,bouhon2014current,Vanderbilt_smooth_gauge, Wieder_axion,Cornfeld_2021, Ninsulator, PhysRevB.92.085126,InvTIVish,afmsplitting, codefects2, semimetals, regdef, Shiozaki14,Jo2020_axion, flatbands2019, Mode2, Weylrmp,UnsupMach, Ahn2019, UnifiedBBc,SchnyderClass}, culminating in versatile classification schemes~\cite{clas3,Wi2,clas4,clas5,BBS_nodal_lines}, most existing approaches by and large trace back to evaluating symmetry representations of bands in the Brillouin zone and are characterized by single gap topological invariants. Recently, a class of topological phases beyond such schemes have been steadily attracting attention{~\cite{Wu1273, bouhon2019nonabelian, BJY_nielsen, Zhao_PT}}. These topological phases instead depend on the wavefunction geometry due to multi-gap conditions~\cite{Wu1273,bouhon2020geometric}. Band degeneracies may then carry non-Abelian topological charges akin to disclinations or vortices in bi-axial nematics~\cite{Nissinen2016,Kamienrmp, Prx2016, volovik2018investigation, Beekman20171}. Braiding these band degeneracies can lead to novel non-trivial multi-gap topological invariants. A prototype example is the Euler class that acts as the paradigmatic real-valued analog of the Chern number~\cite{bouhon2019nonabelian}. Generally, such multi-gap phases can be analyzed using homotopy arguments~\cite{bouhon2020geometric} and are increasingly related to emergent physical effects and retrieved in various experimental contexts~\cite{slager2022floquet,jankowski2023optical,bouhon2023quantum,jankowski2023disorderinduced}. For instance, it was predicted that upon quenching a system with a non-trivial Euler Hamiltonian, monopole-antimonopole pairs can form in the Brillouin zone~\cite{Unal_quenched_Euler}. This phenomenon was recently observed in trapped-ion experiments~\cite{trappedion}. Similarly, Euler class and non-trivial braiding was predicted in phononic systems~\cite{Peng2021,peng2022multi,Park2021,Lange2022} and electronic systems that are strained~\cite{bouhon2019nonabelian,Pspinflip}, undergo a structural phase transition~\cite{chen2021manipulation, magnetic, Lange2021hoti}, or are submitted to an external magnetic field~\cite{guan2021landau}. The most immediate playground for these uncharted topological phases of matter is, however, the  metamaterials~\cite{Guo1Dexp,park2022nodal,Jiang2021,qiu2022minimal, springmass, ezawa2021euler}. The non-Abelian topological charges were recently detected in one-dimensional (1D) electrical circuit metamaterials~\cite{Guo1Dexp}. Meanwhile, in two-dimensional (2D) acoustic metamaterials, the braiding and non-Abelian topological phase transitions were demonstrated~\cite{Jiang2021}. These advancements strongly indicate the {promising} development of this emerging field.

Here, we report the experimental realization of a topological Euler insulator phase in acoustic metamaterials which is featured with an unconventional meron (i.e., half-Skyrmion) configuration in the bulk Bloch states. Remarkably, the meron number of the bulk Bloch bands is connected to the Euler class topological index ($\chi\in \mathbb{Z}$) which is a prototype multi-gap topological invariant. We theoretically show an intricate cancellation of bulk Zak phases in the topological Euler insulator phase that permits an odd Euler class ($\chi=1$) in this work. The cancellation of the Zak phases is then corroborated by the emergence of the in-gap edge states due to orbitals shifted to the unit-cell boundaries, revealing an unconventional bulk-edge correspondence {associated with the odd Euler class topology}. Using acoustic metamaterial realizations and pump-probe techniques, we manage to observe the topological Euler insulator phase by measuring the acoustic meron wave pattern in the Brillouin zone. In addition, the Euler class topology is revealed by the in-gap edge states due to the Zak phase cancellation which are directly observed in our experiment via pump-probe detection at the edge boundaries. It is worth mentioning that in this work, the pump-probe detection is empowered by a spectral decomposition technique for the measured acoustic responses. This method \textcolor{black}{not only enables us to discover the meronic acoustic wave pattern in this work, but also opens a new pathway} in experimental detection of topological invariants by directly measuring the bulk Bloch wave patterns.

\textcolor{black}{It is worth remarking that the observation of skyrmion and meron types of states (as well as their generalizations) in photonic and acoustic systems have inspired lots of research~\cite{skyrmion1,skyrmion2,skyrmion4,meron1,skyrmion5,skyrmion6,skyrmion-review}. Due to the unconventional properties of these states and their potential applications in topologically robust information processing, sensing, and lasing, these states have been studied extensively recently~\cite{skyrmion-review}. However, to date, meronic acoustic waves have not yet been observed in experiments. Our work unveils the first observation of meronic wave patterns in acoustic systems in wavevector space, which sets a benchmark in the study of acoustic states. The unique connection between the meronic wave pattern in wavevector space and the topological Euler insulator phase also uncover intriguing physics that has not yet been revealed before.}

\begin{figure}[t]
\centering
\includegraphics[width=0.96\linewidth]{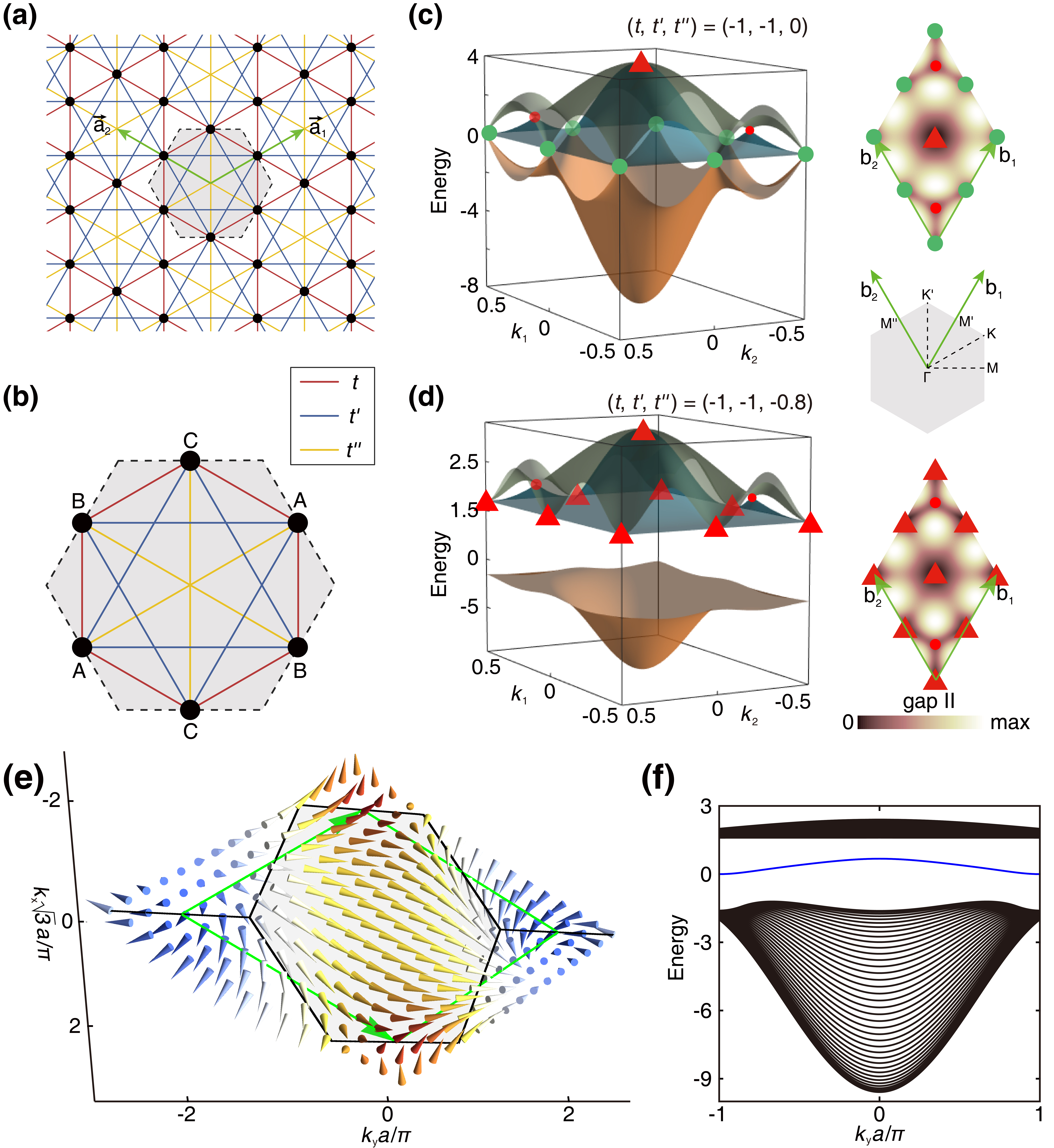}
\caption{Theoretical results. (a) Kagome tight-binding model with unit-cell delineated by the dashed lines. The lattice vectors, \textbf{a}$_1$ and \textbf{a}$_2$, are denoted by green arrows. (b) The unit-cell structure. The NN, NNN, and TNN couplings are denoted as $t$, $t'$ and $t''$, respectively (colored lines). The three inequivalent sublattice sites A, B, and C are labeled by black dots. (c) Bulk dispersion of a topological Euler semimetal with $t=t^{\prime}=-1$ and $t^{\prime\prime}=0$. The right panels show \textcolor{black}{the following: (i) the band degeneracy points between all the three bands (upper) in the rhombus Brillouin zone [linear (Dirac) crossings are represented by circles and quadratic (double) crossings by triangles] where the color map indicates the energy difference between the second and third bands, and (ii) the hexagonal Brillouin zone (lower gray hexagon) with the reciprocal primitive vectors {$\boldsymbol{b}_1$ and $\boldsymbol{b}_2$} (green) chosen as the reference frame for the coordinates $k_1$ and $k_2$ [contrary to (e) that uses the $k_x$ and $k_y$ coordinates].} (d) The bulk dispersion of a topological Euler insulator emerges for finite $t^{\prime\prime}$, here with $t=t^{\prime}=-1$ and $t^{\prime\prime}=-0.8$. \textcolor{black}{The right panel shows the band degeneracy points between the second and third bands in the rhombus Brillouin zone.} (e) Vector distribution in the Brillouin zone (an oblique view) representing the \textcolor{black}{cell-periodic part [in real space]} of the Bloch wavefunction of the first band in the topological Euler insulator phase in (d). Green arrows indicate the {rhombus first} Brillouin zone corresponding to (c) and (d). Black lines indicate the hexagonal {first} Brillouin zone. (f) Energy spectrum of a supercell with the zigzag edge boundaries {where the} edge states in the {bulk} band gap are highlighted by the blue line.} 
\label{Fig.1}
\end{figure}

\sect{Theoretical results}
We consider a three-band tight-binding model in 2D kagome lattice with the NN (nearest neighbor), NNN (next-to-nearest neighbor), and TNN (third nearest neighbor) couplings. There are three inequivalent lattice sites in each unit-cell, labeled separately as A, B, and C. These sites are coupled to each other \textcolor{black}{via} the NN, NNN, and TNN couplings \textcolor{black}{which are} denoted as $t$, $t^{\prime}$ and $t^{\prime\prime}$, respectively [\textcolor{black}{see} Fig. 1(a) and (b)]. \textcolor{black}{We remark that without the TNN coupling, the system is always in various gapless semimetallic phases (as discussed thoroughly in Ref.~\cite{Jiang2021}). It is crucial to introduce the TNN coupling} for the emergence of the topological Euler insulator phase. In the presence of both the inversion, or effectively the $180^\circ$ rotation around the $z$ axis ($C_2$), and the time-reversal symmetries (${\cal T}$), once a proper basis is chosen, the Hamiltonian matrix and \textcolor{black}{all} its eigenvectors are real-valued for \textcolor{black}{arbitrary} wavevector $\vec{k}$ and band \textcolor{black}{index}~\cite{bouhon2019nonabelian}. The explicit Hamiltonian matrix and the Bloch bands (\textcolor{black}{including the} dispersions and wavefunctions) are presented \textcolor{black}{in details} in {the} Supplemental Material~\cite{SM}. 

We show two prototype phases in Fig.~1(c) and (d). Figure 1(c) entails a non-Abelian topological semimetalic phase with $t^{\prime\prime}=0$ where the three bands are interconnected. There are triple points (green dots) as the linear crossing points between the three bands (at the $M$ and equivalent points) as well as two Dirac points (at the $K$ and $K^\prime$ points, labeled by red dots) and a quadratic point (the red triangle) in the partial gap between the second and third bands. Without the TNN couplings, the topological semimetals cannot be gapped, \ie, the three bands remain \textcolor{black}{interconnected} via various nodal points~\cite{Jiang2021}. In contrast, with \textcolor{black}{a} finite TNN coupling $t^{\prime\prime}$, a nontrivial topological Euler gap can be introduced \textcolor{black}{between the first and second bands [see Fig.~1(d)]. Note that the vertical axis of Fig.~1(d) (i.e., the axis for the energy) is not uniform. The scale for the positive energy region is smaller than the scale for the negative energy region, which is designed to show the dispersion and band degeneracy points clearly for the second and third bands. These band degeneracy points are also nontrivial as will be discussed below.}

{The topological Euler gap is characterized by the Euler class topological index $\chi$ which in a three-band system can be expressed as follows~\cite{bouhon2019nonabelian,Unal_quenched_Euler}
\begin{equation}
\chi= \frac{1}{2\pi}\int_{BZ}\dd^2k~\bs{n}\cdot \left(\partial_{k_x} \bs{n} \times \partial_{k_y} \bs{n} \right). \label{eq::Eulerdef}
\end{equation}
Here, the real-valued 3D unit vector $\bs{n}(\bs{k})$ is the (\textcolor{black}{cell-periodic} part of the) Bloch wavefunction for the first band expressed in the $\{A,B,C\}$-sublattice basis.
This band topology has a number of unconventional features: First, the second and the third bands have gapless Wilson loop with nontrivial winding (see Supplemental Material~\cite{SM}). This fragile crystalline topology is actually the generic nature of the split elementary band representation of the kagome lattice~\cite{clas5,Ft1,bouhon2018wilson,bouhon2019nonabelian} \textcolor{black}{(see Supplementary Material~\cite{SM} for more details)}.} Second, in our system the vector {representation of the first bulk band $\bs{n}(\bs{k})$ has an emergent} {\it meron pattern} (i.e., half of a skyrmion), see Fig.~1(e) [the green arrows entail {the} primitive vectors {of the reciprocal lattice}\textcolor{black}{; note that as the arrows here represent the real eigenvectors of the real Bloch Hamiltonian, they have a $\pm1$-gauge degree of freedom. That is, if the vector is reversed, it represents the same eigenstate.}]. In fact, the meron number of {the Bloch eigenvector} is directly connected to the Euler class $\chi=1$ of the topological Euler insulator phase through Eq.~(1) which characterizes the meronic or skyrmionic geometry of the Bloch wavefunction $\bs{n}(\bs{k})$ in the Brillouin zone (see more discussions in {the} Supplemental Material~\cite{SM}, \textcolor{black}{where we show that the integrand of Eq.\;(1) is periodic over one rhombus Brillouin zone, while $\bs{n}(\bs{k})$ is only periodic over four rhombus Brillouin zones within which it forms two Skyrmion windings.}). Third, there are two Dirac points (at $K$) and four quadratic points (at $\Gamma$ and at $M$) connecting the second and the third bands which cannot be completely gapped as long as the $C_2{\cal T}$ symmetry is preserved. This is because these band nodes carry a \textcolor{black}{nontrivial} total patch Euler class 1 that forbids all of them to be annihilated together \textcolor{black}{to open a band gap between the second and third bands}~\cite{BJY_linking,bouhon2019nonabelian}.

{\textcolor{black}{We remark that} he above features are also linked to the bulk-edge response via the Zak phase. \textcolor{black}{For this purpose}, it is crucial {to note} that {in our model} the sublattice sites are sitting at the boundary of a unit-cell, such that the system has inherent Zak phase $\pi$ even in trivial phases ({i.e., the} atomic limit). {More precisely, in the case of a zigzag [referring to the underlying hexagonal lattice] edge termination of the kagome lattice, the perpendicular 1D chain of projected atomic sites starts and ends with atomic sites on the boundary of the 1D unit cell [a feature shared with the honeycomb (graphene) lattice when terminated by the same zigzag edge]. As a consequence and similarly to graphene, the vanishing Zak phase indicates the presence of topological edge states for the zigzag termination}~\footnote{We remind that a Zak phase is a Berry phase computed over a non-contractible path of the Brillouin zone, \ie, a path along a periodic cycle of the Brillouin zone.}. In the topological Euler insulator phase, the $\pi$-Zak {phases along both reciprocal lattice vectors, $\boldsymbol{b}_1$ and $\boldsymbol{b}_2$ [Fig.~1(c)], are} canceled by $\pi$-winding of the Bloch wavefunction due to the meronic pattern in Fig.~1(e)}. We stress that this is {distinct from} the three-band Euler insulator phases considered previously~\cite{bouhon2020geometric,BzduSigristRobust} that are all connected to an {\it orientable} flag atomic limit, leading to the constraint of an {\it even} Euler class that must correspond to an integer multiple of skyrmions instead of merons {of the bulk Bloch wavefunctions in the Brillouin zone}~\cite{bouhon2019nonabelian,Unal_quenched_Euler} {which} shows the {profound} nature from a fundamental perspective as described in the Supplemental Material~\cite{SM}. {Such cancellation leads to the emergence of the edge states in the topological band gap due to the vanishing Zak phase [see Fig.~1(f)]. {Interestingly, in the case of an armchair edge termination, the perpendicular 1D chain of projected atomic sites now starts and ends with the site at the center of the 1D unit cell. This has the consequence to reverse the BBC of the zigzag edge, namely the vanishing Zak phase now indicates the absence of topological edge states. We actually find {\it two} in-gap edge states at the armchair termination which cannot be traced directly from the Zak phase [see the Supplemental Material~\cite{SM}]}. We emphasize that these {zigzag and armchair edge state configurations} are unique features of topological Euler insulators with an odd Euler class that are not found before.}

\begin{figure*}[t!]
\centering\includegraphics[width=\textwidth]{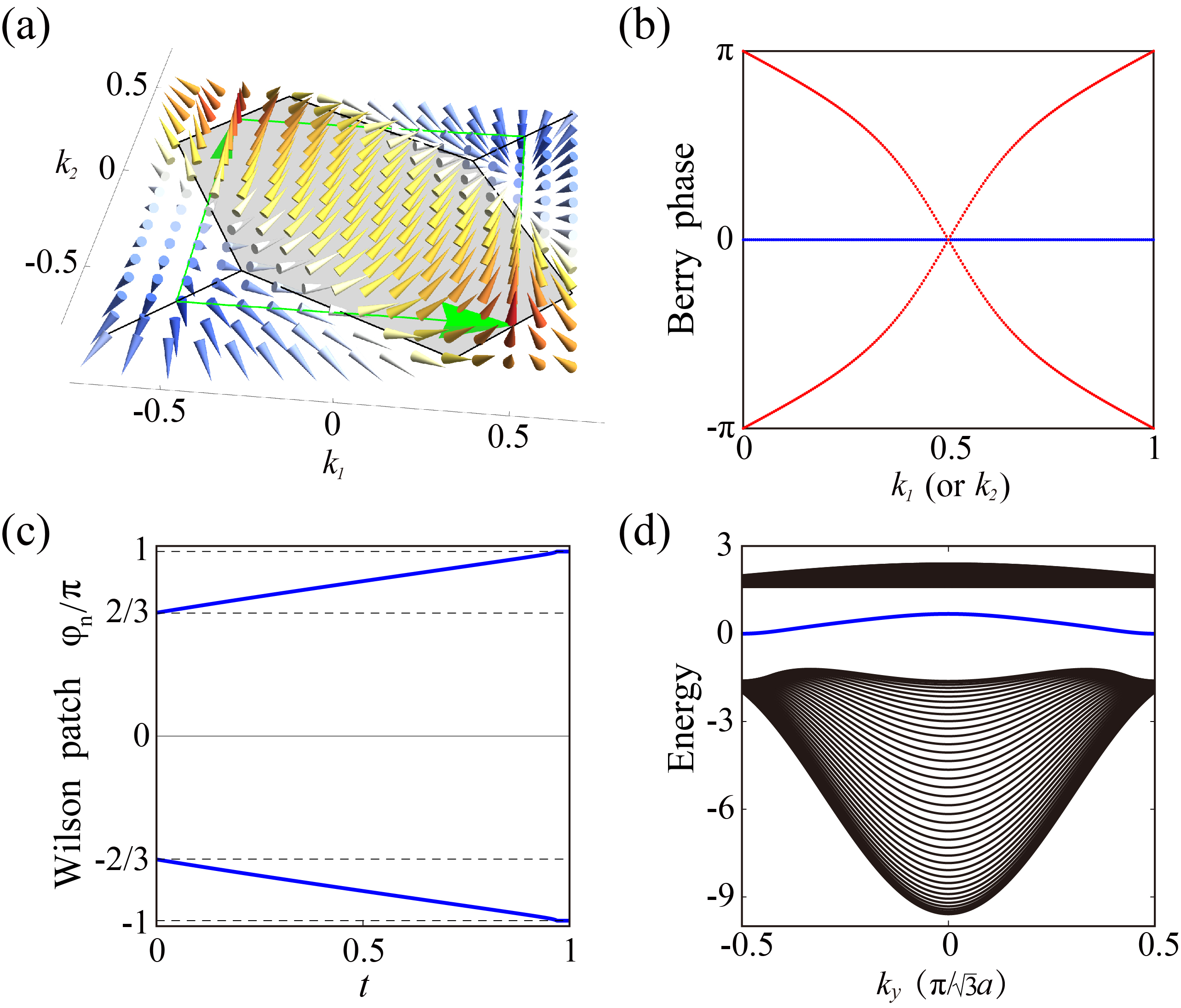}
\caption{Experimental results. (a) Photograph of the {air-borne Euler} acoustic metamaterial. Inset: zoom-in photo of one unit-cell {with the unit-cell boundary indicated by the yellow lines and sublattice sites labeled by the characters. (b) The structure of the air region in a unit-cell (geometric details are given in {the} Supplemental Material) where the unit-cell boundary is indicated by the dashed lines}. (c) Acoustic bulk band structures {obtained} from simulation (yellow curves) and experiments (color map). (d) Measured (color map) and simulated (gray markers) dispersions of the acoustic zigzag edge states. {Gray} dots represent simulated bulk bands projected to the zigzag edge Brillouin zone. (e) Illustration of the pump-probe technique for the measurement of the acoustic Bloch wavefunctions. (f) {The measured acoustic meron pattern, i.e., the vector Bloch wavefunctions of} the first bulk band {in our acoustic metamaterial} {(an oblique view))}. Green arrows indicate {the rhombus Brillouin zone} as in Fig.\;\ref{Fig.1}(e), while {black lines indicate the hexagonal Brillouin zone.} (g)-(h) For the {Bloch} vector {distribution}, the variation of the azimuth and elevation angles along the (g) $M$-$\Gamma$-$M$ and (h) $M$-$K$-$\Gamma$-$K^\prime$-$M$ lines {are presented} {with the} results from both {the} tight-binding model (TBM) and {the} experiments (Exp.).
}
\label{Fig.2}
\end{figure*}

\sect{Materials and methods}
To \textcolor{black}{confirm} the {topological Euler insulator phase} in experiments, we design and fabricate an air-borne acoustic metamaterial, see Fig.~2(a). The unit-cell structure is illustrated in the insets with a lattice constant $a=36\sqrt{3}$~mm. There are three cylindrical acoustic cavities (labeled as A, B, and C), representing three sites in the unit-cell of the tight-binding model in Fig.~1. The NNN and TNN couplings are realized by tubes (with radii $r_1$ and $r_2$, respectively) connecting these cavities. Note that tubes for the NNN couplings are realized by two layers to enforce the inversion symmetry. The tubes representing TNN couplings intersect at the unit-cell center which also provide the NN couplings through indirect {processes which are prominent due to the strong couplings between the acoustic cavities}. By tuning the radii of these tubes, we can realize the topological Euler insulator phase in the acoustic bands. The bulk acoustic bands from both simulation and experiments are presented in Fig.~2(b). The simulation is based on {solving} the acoustic wave equation using commercial finite-element simulation methods (see Supplemental Material~\cite{SM}). The experiments are based on acoustic pump-probe measurements. Specifically, an acoustic source {(a very small speaker)} is placed {in a cavity} at the center of the system (indicated by the blue star) to excite the bulk Bloch acoustic waves. A detector (a small microphone) is used to scan the acoustic wavefunction in the whole system. By varying the excitation and detection frequency (they are always fixed to be equal), we obtain the acoustic wavefunctions at different frequencies. Upon Fourier transformation of the probed wavefunctions, we obtain the dispersions of the excited bulk Bloch acoustic waves. 

\sect{Experimental results} 
\textcolor{black}{As shown in Fig.~2(c), the measured dispersion of the bulk bands agrees quite well with the acoustic bulk band structure obtained from the finite-element simulation. It is encouraging to see that the details of the second and third bands can also be reproduced in the experiments, beside the obvious} band gap between the first band the remaining bands. \textcolor{black}{In fact, there are two Dirac points (at the $K$ points) and four quadratic points (three at the $M$ points and one at $\Gamma$)} {between the second and the third bands}. In the {Supplementary Material}~\cite{SM}, we also give the measured dispersions around these degeneracy points \textcolor{black}{which also confirm the main results from the tight-binding calculations and the finite-element simulations}.

Next we measure the dispersion of the acoustic edge states. By placing the acoustic source at the center of a zigzag edge [the red triangle in Fig.~2(a)], we are able to excite the edge states within the bulk band gap and measure their dispersion using similar pump-probe techniques (i.e., scanning the acoustic wavefunctions along the zigzag edge at various excitation frequencies and then performing the Fourier transformation of the detected {real-space} wavefunction along the edge direction). \textcolor{black}{Figure~2(d) shows that} the measured dispersion {of the edge states} is comparable with the simulated dispersion (the \textcolor{black}{gray markers}), confirming the emergence of the edge states in the {topological Euler} band gap. \textcolor{black}{We note that due to the finite-size effect and the intrinsic dissipation, the valence band edge is blue-shifted and broadened. (Similarly, other bulk states are also shifted and broadened as shown in the figure.) Nevertheless, the measured dispersion of the edge states agree well with the calculated edge dispersion from finite-element simulation of the eigenstates.} {We emphasize again that the topological Euler phase here (as a prototype fragile topological phase protected by the $C_2{\cal T}$ symmetry) does not support robust gapless edge states. The observed edge states are \textcolor{black}{rather} due to the nontrivial Zak phase \textcolor{black}{which is indirectly connected to the Euler topology in kagome lattices}. As explained in details in the Supplemental Material~\cite{SM}, this is a unique phenomenon \textcolor{black}{for} odd Euler class \textcolor{black}{topological phases}. In fact, the Zak phase physics dominates the emergence of the edge states in this work. For instance, we find two branches of edge states for the armchair edge boundaries which stem from the evolution of the edge states in both the complete gap I and the partial gap II due to the Zak phase under the influence of chiral symmetry breaking due to the TNN couplings (see Supplemental Material~~\cite{SM}). That is, one branch of the edge states comes from the partial gap II which is consistent with the observation in Ref.~\cite{Jiang2021}. These phenomena enrich our understanding on the Euler topological phases.}

We now reveal the most salient feature {of the topological Euler gap in our acoustic system} --- the meron pattern {in the bulk Bloch waves}. For this purpose, we develop a technique of spectral decomposition based on the acoustic pump-probe {measurement}. The underlying {principle} is based on the fact that the two-points {acoustic} pump-probe {measurement} gives {the two-points {acoustic} response function which is proportional to the retarded} two-points Green's function of the acoustic waves. At a frequency of excitation and detection $\nu$, the two-points {response} function is a $3\times 3$ tensor since the source and detector can be {allocated} at the A, B, or C sublattice site in different {pump-probe} configurations. Specifically, the response function $\chi_{\alpha\beta}(\vec{r}_s, \vec{r}_d, \nu)$ is a $3\times 3$ tensor where $\alpha, \beta=(A, B, C)$ denote the pumping and detection sites, respectively. $\vec{r}_s$ ($\vec{r}_d$) denotes the position vector of the unit-cell center that the pumping (detection) site belongs to {[see illustration in Fig.~2(e)]}. Upon Fourier transformation in both space and time, the measured response function {becomes a function of wavevector and frequency,} $\chi_{\alpha\beta}(\vec{k}, \nu)$. Theoretically, {the proportionality between the response function and the retarded Green's function of acoustic waves {is}}
\be
\chi_{\alpha\beta}(\vec{k}, \nu) \propto \sum_n \frac{u_{n\vec{k}}^{\alpha~\ast} u_{n\vec{k}}^\beta}{\nu-(\nu_{n\vec{k}}+i\gamma_{n\vec{k}})} ,
\ee
where $n=(1,2,3)$ is the band index, $\nu_{n\vec{k}}$ and $\gamma_{n\vec{k}}$ are{, respectively,} the eigenfrequency and the damping of the Bloch states of the $n$-th band at the wavevector $\vec{k}$. $u_{n\vec{k}}$ is the eigenvector of the Bloch states expressed in the local basis of the sublattice sites A, B, and C. To {determine} the Bloch {eigenvector} $u_{1\vec{k}}$ of the first {acoustic} band, we first {obtain} the {acoustic} response function $\chi_{\alpha\beta}(\vec{k}, \nu)$ through the pump-probe {measurement and the Fourier transformation}. We then examine the response function {at the condition} {when the frequency is} at resonance with the first acoustic {bulk Bloch} band, {i.e.}, $\nu=\nu_{1\vec{k}}$, where the dominant contribution {of the acoustic response function} must come from the first acoustic {bulk} band. We use a singular value decomposition {of the acoustic response function} $\chi_{\alpha\beta}(\vec{k}, \nu_{1\vec{k}})$ to extract this dominant contribution. This process also gives the {Bloch} eigenvector $u_{1\vec{k}}$ as the eigenvector associated with the largest singular value of the {acoustic} response tensor $\chi_{\alpha\beta}(\vec{k}, \nu_{1\vec{k}})$ 
(see Supplemental Material~\cite{SM} for {more} details). By properly tuning the overall phase factor of {the eigenvector} $u_{1\vec{k}}$ {via the redundant gauge degree of freedom}, we can map {the eigenvector into} a real-valued 3D unit vector $\bs{n}(\bs{k})$ {which is then compared with the Bloch wavefunction of the first band in the $\{A,B,C\}$-sublattice basis calculated from the tight-binding theory.} 

{Using this method, we observe for the first time the meron pattern in acoustic waves: Figure~2(f) gives the vector Bloch wavefunctions of the first bulk band in our acoustic metamaterial which show clearly a meron pattern in agreement with Fig.~2(a).} We {further} check quantitatively the azimuthal and elevation angles of the measured eigenvector and the calculated eigenvector along two special lines, the $M$-$\Gamma$-$M$ line [Fig.~2(g)] and the $M$-$K$-$\Gamma$-$K^\prime$-$M$ line [Fig.~2(h)]. The consistency between the {experimental results} and {the tight-binding} theory confirms the nontrivial meronic configuration of the bulk Bloch states and signifies the first observation of acoustic meron {which emerges due to the topological Euler insulator phase here}.

\sect{Conclusion and discussions}
Our experiments unequivocally demonstrate a unique topological Euler insulator phase in an acoustic setup with an unprecedented meronic {pattern}. {Remarkably, the meron topological number is connected to the Euler class topological index in our kagome system. This acoustic meron pattern enriches our understanding on acoustic waves and gives an excellent example of the direct measurement of bulk topological properties. Furthermore, the observed meron pattern} can be generalized to, e.g., skyrmion {patterns} that characterize the non-Abelian topology in systems with more bands {or with an even Euler class}~\cite{braidingtwo}. From {the} experimental {perspective}, the discovery here may inspire future exploration of rich topological {states} with unconventional {Bloch} wavefunction {patterns} and thus opens a new realm for topological physics and {materials}.

\sect{Acknowledgements}
Jian-Hua Jiang thanks the National Key R\&D Program of China (2022YFA1404400), the National Natural Science Foundation of China (Grant Nos. 12125504 and 12074281), the ``Hundred Talents Program'' of the Chinese Academy of Sciences, and the Priority Academic Program Development (PAPD) of Jiangsu Higher Education Institutions. Adrien Bouhon was partially funded by a Marie-Curie fellowship, grant no. 101025315 and acknowledges financial support from the Swedish Research Council (Vetenskapsradet) grant no. 2021-04681. Robert-Jan Slager acknowledges funding from a New Investigator Award, EPSRC grant EP/W00187X/1, a EPSRC ERC underwrite grant  EP/X025829/1, and a Royal Society exchange grant IES/R1/221060, as well as Trinity College, Cambridge.

\sect{Author contributions}
Jian-Hua Jiang, Robert-Jan Slager, and Adrien Bouhon guided the research. Robert-Jan Slager and Adrien Bouhon established the underlying theory. Bin Jiang and Jian-Hua Jiang designed the metamaterial system and the measurement methods. Bin Jiang, Shi-Qiao Wu, Ze-Lin Kong, and Zhi-Kang Lin achieved the experimental observation and data analysis. All authors contributed to the discussion of the results. Jian-Hua Jiang, Robert-Jan Slager, and Adrien Bouhon wrote the main text. 
Bin Jiang, Adrien Bouhon and Robert-Jan Slager wrote the supplementary notes.

\bibliography{reference}

 \clearpage

\includepdf[pages=1]{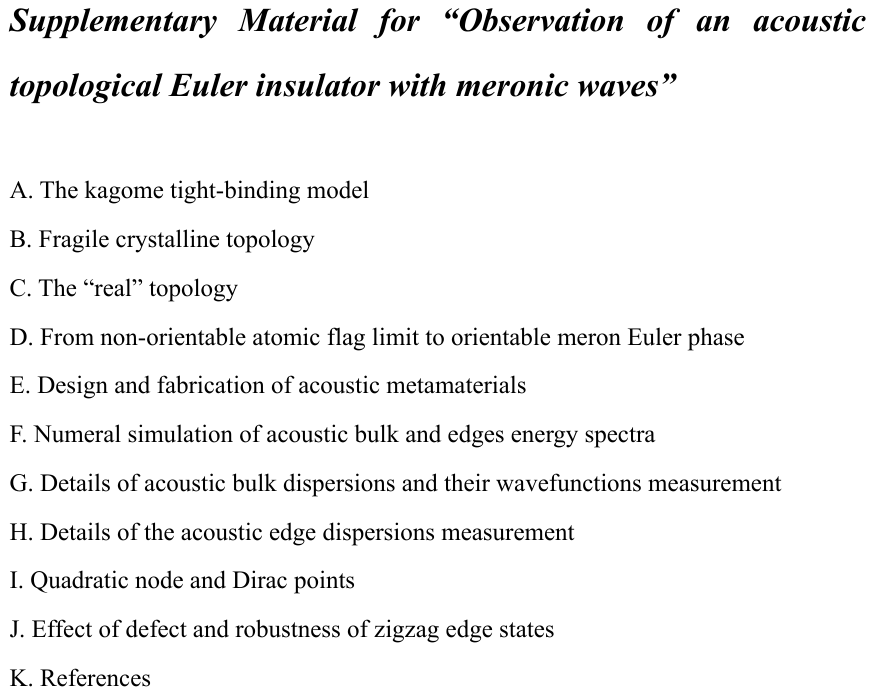}
 \clearpage \newpage
\includepdf[pages=2]{SI.pdf}
 \clearpage \newpage
\includepdf[pages=3]{SI.pdf}
 \clearpage \newpage
\includepdf[pages=4]{SI.pdf}
 \clearpage \newpage
\includepdf[pages=5]{SI.pdf}
 \clearpage \newpage
\includepdf[pages=6]{SI.pdf}
 \clearpage \newpage
 \includepdf[pages=7]{SI.pdf}
 \clearpage \newpage
\includepdf[pages=8]{SI.pdf}
 \clearpage \newpage
\includepdf[pages=9]{SI.pdf}
 \clearpage \newpage
\includepdf[pages=10]{SI.pdf}
 \clearpage \newpage
\includepdf[pages=11]{SI.pdf}
 \clearpage \newpage
\includepdf[pages=12]{SI.pdf}
 \clearpage \newpage
  \includepdf[pages=13]{SI.pdf}
 \clearpage \newpage
\includepdf[pages=14]{SI.pdf}
 \clearpage \newpage
\includepdf[pages=15]{SI.pdf}
 \clearpage \newpage
\includepdf[pages=16]{SI.pdf}
 \clearpage \newpage
\includepdf[pages=17]{SI.pdf}
 \clearpage \newpage
\includepdf[pages=18]{SI.pdf}
 \clearpage \newpage
 \includepdf[pages=19]{SI.pdf}
 \clearpage \newpage
\includepdf[pages=20]{SI.pdf}
 \clearpage \newpage
\includepdf[pages=21]{SI.pdf}
 \clearpage \newpage
\includepdf[pages=22]{SI.pdf}
 \clearpage \newpage
\includepdf[pages=23]{SI.pdf}
 \clearpage \newpage
\includepdf[pages=24]{SI.pdf}
 \clearpage \newpage
\includepdf[pages=25]{SI.pdf}
 \clearpage \newpage
\includepdf[pages=26]{SI.pdf}
 \clearpage \newpage
\includepdf[pages=27]{SI.pdf}
 \clearpage \newpage
\includepdf[pages=28]{SI.pdf}
 \clearpage \newpage
 \includepdf[pages=29]{SI.pdf}
 \clearpage \newpage
\includepdf[pages=31]{SI.pdf}
 \clearpage \newpage
\includepdf[pages=32]{SI.pdf}
 \clearpage \newpage
\includepdf[pages=33]{SI.pdf}
 \clearpage \newpage
\includepdf[pages=34]{SI.pdf}

\end{document}